\documentclass[a4paper,UKenglish]{lipics-v2018}
 
\usepackage{microtype}
\usepackage{xspace,todonotes}
\usepackage[end]{algpseudocode}
\usepackage{booktabs}
\usepackage{footnote}

\title{External memory BWT and LCP computation for sequence collections with applications}
\titlerunning{External memory LCP and BWT computation \textbf{}with applications} 

\author{Lavinia Egidi}{University of Eastern Piedmont, {Alessandria, Italy}}{lavinia.egidi@uniupo.it
}{https://orcid.org/0000-0002-9745-0942}{L.E. was partially supported by the University of Eastern Piedmont project  {\sl Behavioural Types for Dependability Analysis with Bayesian Networks}}
\author{Felipe A. Louza}{Department of Computing and Mathematics, University of São Paulo, {Ribeirão Preto, Brazil}}{louza@usp.br}{https://orcid.org/0000-0003-2931-1470}{F.A.L. was supported by the grant $\#$2017/09105-0 from the São Paulo Research Foundation (FAPESP)}
\author{Giovanni Manzini}{University of Eastern Piedmont, {Alessandria, Italy}\\IIT CNR, Pisa Italy}{giovanni.manzini@uniupo.it}{https://orcid.org/0000-0002-5047-0196}{G.M. was partially supported by PRIN grant 201534HNXC}
\author{Guilherme P. Telles}{Institute of Computing, University of Campinas, {Campinas, Brazil}}{gpt@ic.unicamp.br}{}{G.P.T. acknowledges the support of CNPq}
 
\authorrunning{L. Egidi, F.\,A.\,Louza, G. Manzini, G. P. Telles\textbf{}} 

\Copyright{John Q. Public and Joan R. Public}

\subjclass{Theory of computation $\rightarrow$ Design and analysis of algorithms}


\category{}

\relatedversion{}

\funding{}

\EventEditors{John Q. Open and Joan R. Access}
\EventNoEds{2}
\EventLongTitle{42nd Conference on Very Important Topics (CVIT 2016)}
\EventShortTitle{CVIT 2016}
\EventAcronym{CVIT}
\EventYear{2016}
\EventDate{December 24--27, 2016}
\EventLocation{Little Whinging, United Kingdom}
\EventLogo{}
\SeriesVolume{42}
\ArticleNo{23}
\nolinenumbers 

\begin{document}

\newtheorem{property}[theorem]{Property}

\def\Oh{\mathcal{O}}
\long\def\ignore#1{}
\def\bibignore#1{#1}

\newcommand{\xx}{\$}
\newcommand{\A}{\Sigma}
\newcommand{\Asize}{\sigma}

\newcommand{\locate}{\mathsf{locate}}
\newcommand{\locatep}{\mathsf{locatePrefix}}
\newcommand{\extract}{\mathsf{extract}}
\newcommand{\getint}{\mathsf{getInterval}}

\newcommand{\rank}{\mathsf{rank}}
\newcommand{\sel}{\mathsf{select}}

\newcommand{\lcp}{\ensuremath{\mathsf{lcp}}\xspace}
\newcommand{\lcps}{\mathsf{lcp}}
\newcommand{\lcpzo}{\lcps_{01}}
\newcommand{\lcpz}{\lcps_{0}}
\newcommand{\lcpo}{\lcps_{1}}
\newcommand{\lcpall}{\lcps_{1\cdots k}}

\newcommand{\sort}{\mathsf{sort}}

\newcommand{\Path}{\Pi}
\newcommand{\sa}{\mathsf{sa}}
\newcommand{\sazo}{\mathsf{sa}_{01}}
\newcommand{\SA}{\mathsf{sa}}
\newcommand{\LCP}{\mathsf{LCP}}
\newcommand{\BWT}{\mathsf{BWT}}
\newcommand{\bwt}{\ensuremath{\mathsf{bwt}}\xspace}
\newcommand{\mbwt}{\mathsf{bwt}}
\newcommand{\bwtzo}{\mathsf{bwt}_{01}}
\newcommand{\last}{\mathit{last}}
\newcommand{\Wa}{{W}}
\newcommand{\Wx}{{W}^{-}}
\newcommand{\onex}{\mathbf{1}}
\newcommand{\twox}{\mathbf{2}}
\newcommand{\zerox}{\mathbf{0}}
\newcommand{\LF}{LF}
\newcommand{\Lx}{L}
\newcommand{\Kx}{K}
\newcommand{\tx}{\mathsf{t}}
\newcommand{\Count}{C}
\newcommand{\eps}{$\epsilon$}
\def\stri#1{\mbox{\sf #1}}

\newcommand{\tz}{\mathsf{s}_0}
\newcommand{\tone}{\mathsf{s}_1}
\newcommand{\tzo}{\mathsf{s}_{01}}

\newcommand{\tk}{\mathsf{s}_k}
\newcommand{\tall}{\mathsf{s}_1,\ldots,\mathsf{s}_k}
\newcommand{\tj}{\mathsf{s}_j}
\newcommand{\sumj}[1]{\Sigma_{h=1}^{#1}n_h}

\newcommand{\SAall}{\SA_{1\cdots k}}

\newcommand{\txx}[1]{\mathsf{s}_{#1}}
\newcommand{\nxx}[1]{\mathsf{n}_{#1}}
\newcommand{\tcat}[1]{\txx{#1}^{\mathsf{cat}}}
\newcommand{\sxx}{\mathsf{s}}
\newcommand{\saxx}{\ensuremath{\mathsf{sa}}}
\newcommand{\nz}{{n_0}}
\newcommand{\none}{{n_1}}
\newcommand{\nk}{{n_k}}

\newcommand{\eos}{\$}
\newcommand{\eosz}{\eos_0}
\newcommand{\eosone}{\eos_1}
\newcommand{\eosk}{\eos_k}

\newcommand{\kl}{\lambda}
\newcommand{\klu}{{\kl_{1}}}
\newcommand{\kld}{{\kl_{2}}}
\newcommand{\klb}{\mathbf{\lambda}}
\newcommand{\klub}{{\klb_{1}}}
\newcommand{\kldb}{{\klb_{2}}}

\newcommand{\eosx}[1]{\eos_{#1}}
\newcommand{\bwtz}{\mathsf{bwt}_0}
\newcommand{\bwto}{\mathsf{bwt}_1}
\newcommand{\bwtx}[1]{\mathsf{bwt}_{#1}}
\newcommand{\bwtall}{\mbwt_{1\cdots k}}
\newcommand{\bwtl}{\mbwt_\kl}

\newcommand{\oneb}{{\bf 1}}
\newcommand{\zerob}{{\bf 0}}
\newcommand{\Bid}{\mathsf{Block\_id}}
\newcommand{\bid}{\mathsf{id}}
\newcommand{\avelcp}{\mathsf{avelcp}_{01}}
\newcommand{\avelcpx}{\mathsf{avelcp}}
\newcommand{\maxlcpx}{\mathsf{maxlcp}}
\newcommand{\tlist}{\mathsf{top}}
\newcommand{\rellen}{\mathsf{relevantlengths}}
\newcommand{\lcpstack}{\mathsf{lcpStack}}

\newcommand{\bv}[1]{Z^{(#1)}}
\newcommand{\kh}{{b(h)}}
\newcommand{\sbot}{0}

\newcommand{\hm}{{\sf H\&M}}
\newcommand{\gap}{{\sf Gap}\xspace}
\newcommand{\egap}{{\sf eGap}\xspace}
\newcommand{\semigap}{{\sf semi-gap}\xspace}

\newcommand{\useless}{irrelevant}

\def\gSACA{{\sf gSACA-K}\xspace}
\def\SACA{{\sf SACA-K}\xspace}

\newcommand{\bcrbwt}{{\sf BCR}\xspace}
\newcommand{\bcr}{{\sf BCR+LCP}\xspace}
\newcommand{\extlcp}{{\sf extLCP}\xspace}

\newcommand{\egsa}{{\sf eGSA}\xspace}
\newcommand{\mergebwt}{{\sf Phase 2}\xspace}
\newcommand{\mergelcp}{{\sf Phase 3}\xspace}

\newcommand{\bitb}{{\bf bitB}}
\newcommand{\notset}{{\sf noLCP}\xspace}
\newcommand{\justset}{{\sf newLCP}\xspace}
\newcommand{\set}{{\sf saveLCP}\xspace}
\newcommand{\irr}{{\sf irrelevant}\xspace}

\def\Bx{B_{x}}
\def\Zn{Z^{\mathsf{new}}}
\def\Zo{Z^\mathsf{old}}
\def\da{\ensuremath{\mathsf{da}}\xspace}
\def\xlcp{\ensuremath{\mathsf{xlcp}}\xspace}
\def\Eos{\mathsf{eos}}

\algnewcommand\KwTo{\textbf{to }} 
\algnewcommand\KwAnd{\textbf{and }}
\algnewcommand\KwWrite{\textbf{write }}
\algnewcommand\KwTofile{\textbf{to file }}

\newcommand{\red}{\textcolor{red}}
\newcommand{\blue}{\textcolor{blue}}
\newcommand{\orange}{\textcolor{orange}}

\renewcommand{\tt}{\texttt}

\newcommand{\etal}{{\it et al.}\xspace}

\maketitle

\begin{abstract}
We propose an external memory algorithm for the computation of the BWT and LCP array for a collection of sequences. Our algorithm takes the amount of available memory as an input parameter, and tries to make the best use of it by splitting the input collection into subcollections sufficiently small that it can compute their BWT in RAM using an optimal linear time algorithm. Next, it merges the partial BWTs in external memory and in the process it also computes the LCP values. We prove that our algorithm performs $\Oh(n\, \avelcpx)$ sequential I/Os, where $n$ is the total length of the collection, and $\avelcpx$ is the average Longest Common Prefix of the collection. This bound is an improvement over the known algorithms for the same task. The experimental results show that our algorithm outperforms the current best algorithm for collections of sequences with different lengths and for collections with relatively small average Longest Common Prefix. 

In the second part of the paper, we show that our algorithm can be modified to output two additional arrays that, used with the BWT and LCP arrays, provide simple, scan based, external memory algorithms for three well known problems in bioinformatics: the computation of maximal repeats, the all pairs suffix-prefix overlaps, and the construction of succinct de Bruijn graphs. To our knowledge, there are no other known external memory algorithms for these problems.

\end{abstract}

\section{Introduction}

A fundamental problem in bioinformatics is the ability to efficiently search into the billions of DNA sequences produced by NGS studies. The {Burrows Wheeler transform} (BWT) is a well known structure which is the starting point for the construction of compressed indices for collection of sequences\bibignore{~\cite{books/MBCT2015}}. The BWT is often complemented with the {Longest Common Prefix} (LCP) array since the latter makes it possible to efficiently emulate Suffix Tree algorithms\bibignore{~\cite{GogO13,NM-survey07}}. The construction of such data structures is a challenging problem. Although the final outcome is a {\em compressed} index, construction algorithms can be memory hungry and the necessity of developing {\em lightweight}, i.e. space economical, algorithms was recognized since the
very beginning of the field\bibignore{~\cite{BurKar03,lcp_swat,MF02}}. When even
lightweight algorithms do not fit in RAM, one has to resort to external
memory construction algorithms (see~\cite{latin10j,jea/KarkkainenK16,mics/KarkkainenK17,almob/LouzaTHC17} and references therein).

Although the space efficient computation of the BWT in RAM is well studied, and remarkable advances have been recently obtained~\cite{stoc/Belazzougui14,soda/MunroNN17}, for external memory computation the situation is less satisfactory. For collections of sequences, the first external memory algorithm is the \bcrbwt\ algorithm described in~\cite{tcs/BauerCR13} that computes the multi-string BWT for a collection of total size $n$, performing a number of sequential I/Os proportional to $nK$, where $K$ is the length of the longest sequence in the collection. This approach is clearly not competitive when the sequences have non homogeneous lengths, and it is far from the theoretical optimal even for sequences of equal length. Nevertheless, the simplicity of the algorithm makes it very effective for collections of relatively short sequences, and this has become the reference tool for this problem. This approach was later extended~\cite{jda/CoxGRS16} to compute also the LCP values with the same asymptotic number of I/Os. When computing also the LCP values, or when the input strings have different lengths, the algorithm uses $\Oh(m)$ words of RAM, where $m$ is the number of input sequences. 

In this paper, we present a new external memory algorithm for the computation of the BWT and LCP array for a collection of sequences. Our algorithm takes the amount of available RAM as an input parameter, and tries to make the best use of it by splitting the input into subcollections sufficiently small so that it can compute their BWT in internal memory using an optimal linear time algorithm. Next, it merges the partial BWTs in external memory and in the process it also computes the LCP values. Since the LCP values are computed in a non-standard order, the algorithm is completed by an external memory merge sort procedure that computes the final LCP array. We prove that our algorithm performs $\Oh(n\, \avelcpx)$ sequential I/Os, where $\avelcpx$ is the average Longest Common Prefix of the input sequence. The experimental results show that our algorithm is indeed much faster than \bcrbwt\ for collections of strings of different lengths and when the average LCP is relatively small. 

To our knowledge, the only other known external memory algorithm for computing the BWT and LCP arrays is the one recently proposed in~\cite{corr/BonizzoniVPPR17}. The algorithm performs $\Oh(n\, \maxlcpx)$ sequential I/Os where $\maxlcpx$ is the maximum of the LCP values. Since $\avelcpx \leq \maxlcpx \leq K$ (the maximum string length), in terms of complexity the algorithm in~\cite{corr/BonizzoniVPPR17} is between \bcrbwt\ and our proposal while the RAM usage is $\Oh(m+k)$. We plan to experimentally compare this algorithm to ours in the near future.

Another contribution of the paper, which follows from our main result, is the design of simple external memory algorithms for three well known problems, namely: the computation of maximal repeats~\cite{tcbb/KulekciVX12,spire/OhlebuschGK10}, the computation of the all pairs suffix-prefix overlaps~\cite{ipl/GusfieldLS92,ipl/OhlebuschG10,jda/TustumiGTL16}, and the construction of succinct de Bruijn graphs~\cite{latin/BelazzouguiGMPP16,dcc/BoucherBGPS15,wabi/BoweOSS12}. This is achieved using the BWT and LCP arrays, together with two additional arrays that our algorithm can compute without any asymptotic slowdown. The first one is the so called Document Array providing for each suffix the ID of the sequence it belongs to; the second one is a bit array indicating whether each suffix is a substring of the one immediately following it in lexicographic order. Our external memory algorithms for these problems are derived from known internal memory algorithms, but they process the input data in a single sequential scan. In addition, for the problem of the all pairs suffix-prefix, we go beyond the recent solutions~\cite{ipl/OhlebuschG10,jda/TustumiGTL16} that compute {\em all} the overlaps, by computing only the overlaps above a certain length, still spending constant time per reported overlap.

\ignore{In the next section we fix the notation that we use in the rest of the paper and we briefly review the algorithms on which we base our proposal. Section~\ref{sec:algorithm} presents our algorithm and discusses its costs. In Section~\ref{sec:exp} we report the results of our experiments. Section~\ref{sec:apps} is devoted to the applications.}

\section{Background}\label{sec:notation}\label{sec:bblocks}

Let $\txx{}[1,n]$ denote a string of length $n$ over an alphabet $\A$ of
size~$\sigma$. As usual, we assume $\txx{}[n]$ is a special symbol (end-marker) not appearing
elsewhere in $\txx{}$ and lexicographically smaller than any other symbol. We
write $\txx{}[i,j]$ to denote the substring $\txx{}[i] \txx{}[i+1] \cdots
\txx{}[j]$. If $j\geq n$ we assume $\txx{}[i,j] = \txx{}[i,n]$. If $i>j$ or
$i > n$ then $\txx{}[i,j]$ is the empty string. Given two strings $\txx{1}$
and $\txx{2}$ we write $\txx{1} \preceq \txx{2}$ ($\txx{1} \prec \txx{2}$) to denote
that $\txx{1}$ is lexicographically (strictly) smaller than $\txx{2}$. We denote
by $\LCP(\txx{1},\txx{2})$ the length of the longest common prefix between
$\txx{1}$ and $\txx{2}$.

The {\em suffix array} $\saxx[1,n]$ associated to $\txx{}$ is the permutation
of $[1,n]$ giving the lexicographic order of $\txx{}$'s suffixes, that is,
for $i=1,\ldots,n-1$, $\txx{}[\saxx[i],n] \prec \txx{}[\saxx[i+1],n]$.

The {\em longest common prefix} array $\lcp[1,n+1]$ is defined for $i=2,\ldots,n$
by
\begin{equation}\label{eq:lcpdef}
\lcp[i]=\LCP(\txx{}[\saxx[i-1],n],\txx{}[\saxx[i],n]);
\end{equation}
the $\lcp$ array stores the length of the longest common prefix (LCP) between
lexicographically consecutive suffixes. For convenience we define
$\lcp[1]=\lcp[n+1] = -1$. 

\begin{figure}[t]
\def\xy{$\eosone$}
\def\xz{$\eosx{2}$}
\def\idy{1}
\def\idz{2}
\setlength{\tabcolsep}{5pt}
\begin{center}\sf
\begin{tabular}[t]{ccc}
\begin{tabular}[t]{|r|c|l|}\hline
lcp&bwt& {\em context}\\\hline
 -1 & b & \xy       \\
  0 & c & ab\xy     \\
  2 &\xy& abcab\xy  \\
  0 & a & b\xy      \\
  1 & a & bcab\xy   \\
  0 & b & cab\xy    \\
 -1 &   & \\\hline
\end{tabular}&
\begin{tabular}[t]{|r|c|l|}\hline
lcp&bwt& {\em context}\\\hline
 -1 & c & \xz       \\
  0 &\xz& aabcabc\xz\\
  1 & c & abc\xz    \\
  3 & a & abcabc\xz \\
  0 & a & bc\xz     \\
  2 & a & bcabc\xz  \\
  0 & b & c\xz      \\
  1 & b & cabc\xz   \\
 -1 &   & \\\hline
\end{tabular}&
\begin{tabular}[t]{|r|r|c|l|}\hline
id &$\lcp_{12}$&$\mbwt_{12}$& {\em context}\\\hline
 \idy & -\idy & b & \xy\\
 \idz &  0 & c & \xz       \\
 \idz &  0 &\xz& aabcabc\xz\\
 \idy &  \idy & c & ab\xy\\
 \idz &  2 & c & abc\xz    \\
 \idy &  3 &\xy& abcab\xy\\
 \idz &  5 & a & abcabc\xz \\
 \idy &  0 & a & b\xy\\
 \idz &  1 & a & bc\xz     \\
 \idy &  2 & a & bcab\xy\\
 \idz &  4 & a & bcabc\xz  \\
 \idz &  0 & b & c\xz      \\
 \idy &  1 & b & cab\xy\\
 \idz &  3 & b & cabc\xz   \\
   & -1 &   & \\\hline
\end{tabular}
\end{tabular}
\end{center}
\caption{LCP array and BWT for $\txx{1}=\stri{abcab\xy}$
and $\txx{2}=\stri{aabcabc\xz}$,
and multi-string BWT and corresponding LCP array for the
same strings. Column {\sf id} shows, for each entry of
$\mbwt_{12} = \stri{bc\xz cc\xy aaaabbb}$ whether it comes
from $\txx{1}$ or $\txx{2}$.}\label{fig:BWTetc}
\end{figure}


Let $\tone[1,\none],\ldots,\tk[1,\nk]$ be such that $\tone[\none] = \eosone,\ldots,\tk[\nk]=\eosk$, where 
where $\eosone < \ldots < \eosk$ are $k$ symbols not appearing elsewhere in $\tall$ and smaller than any other symbol. Let $\SAall[1,n]$ denote
the suffix array of the concatenation $\tone\cdots\tk$ of total length $n=\sumj{k}$. The {\em multi-string} BWT~\cite{jda/CoxGRS16,tcs/MantaciRRS07}
of $\tall$, denoted by $\bwtall[1,n]$, is defined as
\begin{equation}\label{eq:mbwtdef}
\bwtall[i] =
\begin{cases}
\tj[\nxx{j}]                      & \mbox{if } \SAall[i] = \sumj{j-1} + 1\\
\tj[\SAall[i]-\sumj{j-1} - 1] & \mbox{if } \sumj{j-1}+1 < \SAall[i] \leq \sumj{j}.
\end{cases}
\end{equation}

Essentially $\bwtall$ is a permutation of the symbols in $\tall$ such that the position in $\bwtall$ of $\txx{i}[j]$ is given by the lexicographic rank of its context $\txx{i}[j+1,n_i]$ (or $\txx{i}[1,n_i]$ if $j=n_i$).
\ignore{In other words, $\bwtall[i]$ is the symbol preceding the $i$-th
lexicographically larger suffix, with the exception that if $\SAall[i] = \sumj{j-1} + 1$ then
$\bwtall[i] = \eosx{j}$.
Hence, $\bwtall[i]$ always comes from the
string $\tj$ ($j\in\{1,\ldots,k\}$) containing the $i$-th largest suffix.} Fig.~\ref{fig:BWTetc} shows an example with $k=2$. Notice that for $k=1$, this is the usual Burrows-Wheeler transform \cite{BW94}.

Given the suffix array $\SAall[1,n]$ of the concatenation $\tone\cdots\tk$, we consider the corresponding LCP array $\lcpall[1,n]$ defined as in~\eqref{eq:lcpdef} (see again Fig.~\ref{fig:BWTetc}). Note that, for $i=2,\ldots,n$, $\lcpall[i]$
gives the length of the longest common prefix between the contexts of
$\bwtall[i]$ and $\bwtall[i-1]$. We stress that all practical implementations use a single $\eos$ symbol as end-marker for all strings to avoid alphabet explosion, but end-markers from different strings are then sorted as described, i.e., on the basis of the index of the strings they belong to.



\medbreak\noindent\textbf{\textsf{Computing {multi-string} BWTs.}} The \gSACA algorithm~\cite{tcs/LouzaGT17}, based on algorithm \SACA~\cite{tois/Nong13}, computes the suffix array for a string collection.
Given a collection of strings of total length $n$, \gSACA computes the suffix array for their concatenation in $O(n)$~time using $(\sigma+1) \log n$ additional bits (in practice, only 2KB are used for \textsf{ASCII} alphabets).  It is optimal for alphabets of constant size $\sigma=O(1)$. The {\em multi-string} $\bwtall$ of $\tall$ can be easily obtained from the suffix array as in~\eqref{eq:mbwtdef}. \gSACA can compute also the \lcp array $\lcpall$ still in linear time using only the additional space for the \lcp values.

\medbreak\noindent\textbf{\textsf{Merging multi-string BWTs.}} The \gap algorithm~\cite{spire/EgidiM17}, based on an earlier algorithm by  Holt and McMillan~\cite{bioinformatics/HoltM14}, is a simple procedure to merge multi-string BWTs. In its original formulation the \gap algorithm can also merge LCP arrays, but in this paper we compute LCP values using a different approach more suitable for external memory execution. We describe here only  the main idea behind \gap and refer the reader to~\cite{spire/EgidiM17} for further details.

Given $k$ multi-string BWTs for disjoint subcollections, the \gap algorithm computes a multi-string BWT for the whole collection. The computation does not explicitly need
the collection but only the multi-string BWTs to be merged. For
simplicity in the following we assume we are merging $k$
single-string BWTs $\bwto=\bwt(\tone),\ldots,\bwtx{k}=\bwt(\tk)$; the algorithm
does not change in the general case where the inputs are multi-string BWTs.
Recall that computing $\bwtall$ amounts to sorting the symbols of $\bwto,\ldots,\bwtx{k}$
according to the lexicographic order of their contexts, where the context of
symbol $\bwtx{j}[i]$ is $\txx{j}[\sa_j[i],n_j]$, for $j=1,\ldots,k$.

The \gap\ algorithm works in successive iterations. After the $h$-th iteration the entries of each $\bwtl$ are sorted on the basis of the first $h$ symbols of their context. More formally, the output of the $h$-th iteration is a $k$-valued vector $\bv{h}$ containing $n_\kl=|\txx{\kl}|$ entries $\klb$ for each $\kl=1,\ldots,k$, such that the following property holds.

\begin{property}\label{prop:hblock}
For $\klu,\kld\in\{1,\ldots,k\}$, $\klu < \kld$, 
and $i=1,\ldots, n_\klu$ 
and $j=1,\ldots, n_\kld$ the $i$-th $\klu$ precedes the
$j$-th $\kld$ in $\bv{h}$ iff $\txx{\klu}[\sa_\klu[i], \sa_\klu[i] + h -1] \;\preceq\; \txx{\kld}[\sa_\kld[j], \sa_\kld[j] + h -1]$.\qed
\end{property}

Following Property~\ref{prop:hblock} we identify the $i$-th $\kl$ in
$\bv{h}$ with $\bwtl[i]$ so
that  $\bv{h}$ corresponds to a permutation of $\bwtall$.
Property~\ref{prop:hblock} is equivalent to state that we can logically
partition $\bv{h}$ into $\kh+1$ blocks
\begin{equation}\label{eq:Zblocks}
\bv{h}[1,\ell_1],\; \bv{h}[\ell_1+1, \ell_2],\; \ldots,\;
\bv{h}[\ell_\kh+1,n]
\end{equation}
such that each block is either a singleton or corresponds to the set of $\bwtall$ symbols whose contexts are prefixed by the same length-$h$ string. Within
each block, for $\klu<\kld$, the symbols of $\bwtx{\klu}$ precede those of $\bwtx{\kld}$ and the context
of any symbol in block $\bv{h}[\ell_j+1, \ell_{j+1}]$ is lexicographically
smaller than the context of any symbol in block $\bv{h}[\ell_k+1,
\ell_{k+1}]$ with $k>j$. We keep explicit track of such blocks using a bit array $B[1,n+1]$ such that at the end of iteration $h$ it is $B[i]\neq 0$ if and only if a block of $\bv{h}$ starts at position~$i$, i.e. $\lcpall[i]=h-1$.
By Property~\ref{prop:hblock}, when all entries in $B$ are nonzero, $\bv{h}$ describes how the $\bwtx{j}$ ($j=1,\ldots,k$) should be merged to get $\bwtall$.

\section{The \egap\ algorithm}\label{sec:algorithm}

At a glance, the \egap\ algorithm for computing the {multi-string} BWT and LCP array in external memory works in three phases.
First it builds multi-string BWTs for sub-collections in internal memory, then it merges these BWTs in external memory and generates the LCP values. Finally, it merges the LCP values in external memory.

\medbreak\noindent\textbf{\textsf{Phase 1: BWT computation.}}
\textbf{}Given a collection of sequences $\txx{1}, \txx{2}, \ldots, \txx{k}$, we split it into sub-collections sufficiently small that we can compute the multi-string SA for each one of them using the linear time internal memory \gSACA\ algorithm (Section~\ref{sec:bblocks}). 
After computing each SA, Phase 1 writes each multi-string BWT to disk in uncompressed form using one byte per character.  


\medbreak\noindent\textbf{\textsf{Phase 2: BWT merging and LCP computation.}}
This part of the algorithm is based on the \gap algorithm described in Section~\ref{sec:bblocks} but it is designed to work efficiently in external memory and it computes LCP values in addition to merging the input (multi-string) BWTs.
In the following we assume that the input consists of $k$ BWTs $\bwt_1,\ldots,\bwt_k$ of total length $n$ over an alphabet of size $\sigma$. The input BWTs are read from disk and never moved to internal memory. We denote by $\bwtall$ and $\lcpall$ the output BWT and LCP arrays.  

The algorithm initially sets $\bv{0} = \onex^{\none}\twox^{n_2}\ldots \mathbf{k}^{n_k}$ and 
$B = \onex \zerox^{n-1} \onex$. Since the context of every symbol is prefixed by the same length-0 string (the empty string), initially there is a single block containing all symbols. At iteration $h$ the algorithm computes $\bv{h}$ from $\bv{h-1}$ as follows. We define an array $F[1,\sigma]$ such that $F[c]$ contains the number of occurrences of characters smaller than $c$ in $\bwtall$. $F$ partitions $\bv{h}$ into $\sigma$ buckets, one for each symbol. Using $\bv{h-1}$ we scan the partially merged BWT, and whenever we encounter the BWT character $c$ coming from $\bwt_i$, with $i\in\{1,\ldots,k\}$, we store it in the next free position of bucket~$c$ in $\bv{h}$; note that $c$ is not actually moved, instead we write $i$ in its corresponding position in $\bv{h}$. Instead of using distinct arrays $\bv{0}, \bv{1}, \ldots $ we only use two arrays $\Zo$ and $\Zn$, that are kept on disk. At the beginning of iteration $h$ it is $\Zo = \bv{h-1}$ and $\Zn = \bv{h-2}$; at the end $\Zn = \bv{h}$ and the roles of the two files are swapped. While $\Zo$ is accessed sequentially, $\Zn$ is updated sequentially within each bucket, that is within each set of positions corresponding to a given character. Since the boundary of each bucket is known in advance we logically split the $\Zn$ file in buckets and write to each one sequentially. 

The key to the computation of the LCP array by \egap\ is to exploit the bitvector $B$ used by \gap to mark the beginning of blocks. We observe that entry $B[i]$ is set to $\onex$ during iteration $h=\lcpall[i]+1$, when it is determined that the contexts of $\bwtall[i]$ and $\bwtall[i-1]$ have a common prefix of length exactly $h-1$ (and a new block is created). 
We introduce an additional bit array $\Bx$ such that, at the beginning of iteration $h$, $\Bx[i]=\onex$ iff $B[i]$ has been set to $\onex$ at iteration $h-2$ or earlier. During iteration $h$, if $B[i]=\onex$ we look at $\Bx[i]$. If $\Bx[i]=\zerox$ then $B[i]$ has been set at iteration $h-1$: thus we output to a temporary file $F_{h-2}$ the pair $\langle i,h-2\rangle$ to record that $\lcpall[i]=h-2$, then we set $\Bx[i]=\onex$ so no pair for position $i$ will be produced in the following iterations. At the end of iteration $h$, file $F_{h-2}$ contains all pairs $\langle i,\lcpall[i]\rangle$ with $\lcp[i]=h-2$; the pairs are written in increasing order of their first component, since $B$ and $\Bx$ are scanned sequentially. These temporary files will be merged in Phase~3.
 
\ignore{Since $B[i]$ is read at the same time as $\Zo[i]$, and $B[j]$ is written at the same time as $\Zn[j]$, we found it convenient to use a reading copy of bitarray $B$ which is coupled with $\Zo$ and a writing copy coupled with $\Zn$. Precisely, the files for $\Zo$ and $\Zn$ consist of $n$ bytes each; one bit of each byte is reserved for $B$.}

 \ignore{The crucial observation is that if two BWT characters in the same bucket come from different blocks they correspond to contexts that differ in the first $h$ characters: when this happens the bit array $B$ is updated to keep track of the creation of a new block. }
As proven in~\cite[Lemma~7]{spire/EgidiM17}, if at iteration $h$ of the \gap algorithm we set $B[i]=\onex$, then at any iteration $g \geq h+2$ processing the entry $\bv{g}[i]$ will not change the arrays $\bv{g+1}$ and $B$. Since the roles of the $\Zo$ and $\Zn$ files are swapped at each iteration,
and at iteration $h$
we scan $\Zo = \bv{h-1}$ to update $\Zn$ from $\bv{h-2}$ to $\bv{h}$, we can compute only the entries $\bv{h}[j]$ that are different from $\bv{h-2}[j]$.
In particular, any range $[\ell,m]$ such that $\Bx[\ell] = \Bx[\ell+1] = \cdots = \Bx[m] = \onex$ can be added to a set of {\em irrelevant} ranges that the algorithm may skip in successive iterations (irrelevant ranges are defined in terms of the array $\Bx$ as opposed to the array $B$, since before skipping an irrelevant range we need to update both $\Zo$ and $\Zn$). We read from one file the ranges to be skipped at the current iteration and simultaneously write to another file the ranges to be skipped at the next iteration (note that irrelevant ranges are created and consumed sequentially). Since skipping a single irrelevant range takes $\Oh(k+\sigma)$ time, an irrelevant range is stored only if its size is larger than a given threshold $t$ and we merge consecutive irrelevant ranges whenever possible. In our experiments we used $t=\max(256,k+\sigma)$. In the worst case the space for storing irrelevant ranges could be $\Oh(n)$ but in actual experiments it was always less than $0.1 n$ bytes.

As in the \gap algorithm, when all entries in $B$ are nonzero, $\Zo$ describes how the BWTs $\bwtx{j}$ ($j=1,\ldots,k$) should be merged to get $\bwtall$, and a final sequential scan of the input BWTs along with $\Zo$ allows to write $\bwtall$ to disk, in sequential order. 
Our implementation can merge at most $2^7 = 128$ BWTs at a time because we use $7$ bits to store each entry of $\Zo$ and $\Zn$. These arrays are maintained on disk in two separate files; the additional bit of each byte are used to keep the current and the next copy of $B$. The bit array $\Bx$ is stored separately in a file of size $n/8$ bytes. To merge of a set of $k>128$ BWTs  we split the input in subsets of cardinality $128$ and merge them in successive rounds. We have also implemented a semi-external version of the merge algorithm that uses $n$ bytes of RAM. The $i$-th byte is used to store $\Zo[i]$ and $\Zn[i]$ (3 bits each), $B[i]$ and $\Bx[i]$.


\ignore{If the input consists of $k>2$ BWTs we need to replace the binary arrays $\bv{h}$ with arrays with entries of size $\lceil\log_2 k \rceil$ bits but no other changes are necessary to the algorithm. An alternative approach is to use a recursive pairwise merging procedure that merges groups of $2, 4, 8, \ldots, 2^{\lceil\log_2 k \rceil - 1}$ BWTs and produces the final merged BWT in $\lceil\log_2 k \rceil$ rounds. Note that the merging of the smaller groups at each round are independent and can be performed in parallel.}

\ignore{We use the bit-array $\Bx$ also for reducing the complexity to  $\Oh( n\, \avelcpx)$, where $\avelcpx$ is the arithmetic average of the entries in the final LCP array (in~\protect{\cite{spire/EgidiM17}} we use instead the concept of monochrome blocks which are not useful in an external memory setting). The idea is to skip the portions of $\bv{h-1}$ for which BWT and LCP values have been already determined.}

\ignore{Note that at each iteration of the (modified) \gap algorithm all data is accessed sequentially. In particular, the input BWTs and the arrays $\Zo$, $B$ and $\Bx$ are scanned sequentially, from beginning to end with the exception of the ranges that are skipped. The arrays $\Zn$ and $B$ are updated sequentially within each bucket, that is within each set of positions corresponding to a given character. Since the boundary of each bucket is known in advance we logically split $\Zn$ and $B$ in buckets and write to each one sequentially with the exception of portions which are skipped when the algorithm reaches an irrelevant range. 

To transform the algorithm to work in external memory we simply need to store the main arrays on disk.}

\medbreak\noindent\textbf{\textsf{Phase 3: LCP merging}}
At the end of Phase~2 all $\LCP$-values have been written to the temporary files $F_h$ on disk as pairs $\langle i,\lcp[i]\rangle$. Each file $F_h$ contains all pairs with second component equal to $h$ in order of increasing first component. The computation of the LCP array is completed using a standard external memory multiway merge\bibignore{~\cite[ch.~5.4.1]{knuth3}} of $\maxlcpx$ sorted files, where $\maxlcpx=\max_i(\lcpall[i])$ is the largest LCP value.

\subsection{Analysis}

During Phase~1, \gSACA\ computes the suffix array for a sub-collection of total length $m$ using $9m$ bytes. If the available RAM is $M$, the input is split into subcollections of size $\approx M/9$. Since \gSACA\ runs in linear time, if the input collection has total size $n$, Phase~1 takes $\Oh(n)$ time overall. 


A single iteration of Phase 2 consists of a complete scan of $\bv{h-1}$ except for the irrelevant ranges. Since the algorithm requires $\maxlcpx$ iterations, without skipping the irrelevant ranges the algorithm would require $\maxlcpx$ sequential scans of $\Oh(n)$ bytes of data. Reasoning as in~\cite[Theorem~8]{spire/EgidiM17} we get that by skipping irrelevant ranges the overall cost of this phase is $\Oh( n\, \avelcpx)$ time and sequential I/Os, where $\avelcpx$ is the arithmetic average of the entries in the final LCP array.

\ignore{

As we detailed earlier, Phase 2 needs two $n$ byte files for $\Zo$, $\Zn$ and the two copies of $B$. 

In the $i$-th byte of the first file we use 7 bits to store $\Zo[i]$ and one additional bit to store $B[i]$; in the second file we store $\Zn$ and an additional copy of the $B$ array. Having two copies of $B$ simplifies the implementation since one is used for reading (the one stored with $\Zo$), and the other for writing (the one stored with $\Zn$): the rationale is that the algorithm reads entry $B[i]$ simultaneously to $\Zo[i]$ and writes entry $B[j]$ simultaneously to the writing of $\Zn[j]$. 
Bitarray $\Bx$ is stored in a dedicated file of size $n/8$ bytes. In addition 
Phase 2 uses at worst $9n$ bytes for the pairs $\langle i,\lcp[i]\rangle$, $5$ for the position $i$ and at most $4$ for the LCP value\footnote{This space usage can be optimized to at most $5n$ bytes if we memorize only the positions $i$ in each $F_h$ file and only the LCP values in the final merge.}.

Since storing an irrelevant range takes $2+k+\sigma$ words and skipping it takes $\Oh(k+\sigma)$ time, an irrelevant range is stored only if its size is larger than a given threshold $t$ and we merge consecutive irrelevant ranges as possible. Our experiments were carried out with a value $t=\max(256,k+\sigma)$. In the worst case the space for storing irrelevant ranges could be $\Oh(n)$ but in actual experiments it was always less than $0.1 n$ bytes.
Therefore the peak disk space usage of our algorithm is less than $12n$ bytes.}

Phase 3 takes $\Oh(\lceil \log_K \maxlcpx\rceil)$ rounds; each round merges $K$ LCP files by sequentially reading and writing $\Oh(n)$ bytes of data. The overall cost of Phase 3 is therefore $\Oh(n \log_K \maxlcpx)$ sequential I/Os. In our experiments we used $K=256$; since in our tests $\maxlcpx < 2^{16}$ two merging rounds were always sufficient.

The above analysis suggests that Phase 2 is the most expensive phase of the \egap algorithm. Indeed, in our experiments we found that Phase 2 always took at least 95\% of the overall running time.


\section{Experiments}\label{sec:exp}

In this section we report some preliminary experiments on the \egap algorithm. Testing external memory algorithms is extremely time consuming since, to make a realistic external memory setting, one has to use an amount of RAM smaller than the size of the data. If more RAM is available, even if the algorithm is supposedly not using it, the operative system will use it to temporary store disk data and the algorithm will be no longer really working in external memory. This phenomenon will be apparent also from our experiments. For this reasons we used datasets of size 8GB, reported in Table~\ref{t:datasets}, and a machine with 32GB of RAM but reduced at boot time to 1GB, to simulate input data much larger than the available RAM, and 8GB, to simulate input data of approximately the same size as the available RAM and test also the semi external version of our algorithm. Note that for a 8GB input, the output BWT+LCP data has size 24GB, so even 8GB RAM is still significantly less than the input and the output combined.

We implemented \egap in {\sf ANSI C} based on the source code of \gap~\cite{spire/EgidiM17} and \gSACA~\cite{tcs/LouzaGT17}. Our algorithm was compiled with \textsf{GNU GCC ver. 4.6.3}, with optimizing option \textsf{-O3}. The source code is freely available at \url{https://github.com/felipelouza/egap/}. The experiments were conducted on a machine with GNU/Linux Debian 7.0/64 bits operating system using an {\sf Intel i7-3770 3.4 GHz} processor with 8 MB cache, 32 GB of RAM and a 2.0 TB SATA hard disk with 7200 RPM and 64 MB cache.

\begin{savenotes}
\begin{table}[!htb]
\centering
\caption{Datasets used in our experiments. 
\textsf{shortreads} are DNA reads from human genome\footnote{\url{ftp://ftp.sra.ebi.ac.uk/vol1/ERA015/ERA015743/srf/}} trimmed to length $100$.
\textsf{longreads} are Illumina HiSeq 4000 paired-end RNA-seq reads from plant Setaria viridis\footnote{\url{https://trace.ncbi.nlm.nih.gov/Traces/sra/?run=ERR1942989}} trimmed to length $300$.
\textsf{pacbio} are PacBio RS II reads from Triticum aestivum (wheat) genome\footnote{\url{https://trace.ncbi.nlm.nih.gov/Traces/sra/?run=SRR5816161}} with different lengths.
\textsf{pacbio.1000} are the strings from \textsf{pacbio} trimmed to length 1,000.
Columns 5 and 6 show the maximum and average lengths of the single strings. 
Columns 7 and 8 show the maximum and average LCPs of the collections. 
}
\label{t:datasets}
\begin{tabular}{@{}|l|c|r|r|rr|rr|@{}}
\hline
Name & Size GB & $\sigma$ & N. of strings & Max Len & Ave Len & Max LCP & Ave LCP \\ \hline
{shortreads}     & 8.0     & 6 & 85,899,345 & 100     & 100     & 99      & 27.90   \\
{longreads}      & 8.0     & 5 & 28,633,115 & 300     & 300     & 299     & 90.28   \\
{pacbio.1000}    & 8.0     & 5 & 8,589,934  & 1,000   & 1,000   & 876     & 18.05   \\
{pacbio}         & 8.0     & 5 & 942,248    & 71,561  & 9,116   & 3,084   & 18.32   \\ \hline
\end{tabular}
\end{table}
\end{savenotes}

We compared \egap with \bcr\footnote{\url{https://github.com/BEETL/BEETL/}} from the BEETL Library~\cite{tcs/BauerCR13} which is the currently most used tool for the construction of BWT and LCP arrays in external memory. The tool \bcr\ cannot handle input sequences of different lengths, so for the {\sf pacbio} dataset we used the tool \extlcp~\cite{jda/CoxGRS16} by the same authors. As a reference we also tested the external memory tool \egsa~\cite{almob/LouzaTHC17} that  computes the Suffix and LCP arrays for a collection of sequences. However, we tested \egsa\ only using 32GB of RAM since the authors in~\cite{almob/LouzaTHC17} showed its running time degrades about 25 times when the RAM is restricted to the input size.

The results in Table~\ref{t:time} show that, as predicted by the theory, \egap's running time per input byte is roughly proportional to the average LCP. Again according to the theoretical analysis in~\cite{tcs/BauerCR13}, \bcr\ running time per input byte is proportional to the sequence length. Using 1GB RAM the tool \bcr\ could not handle the {\sf shortreads} dataset because of insufficient internal memory, and it stopped with an internal error after four days of computation on the {\sf pacbio.1000} dataset. Due to lack of time we could not complete the experiments with \extlcp\ with the {\sf pacbio} dataset. We have contacted the authors about the error and plan to complete the comparison in the near future. We also plan to include in the comparison the two recent algorithms proposed in~\cite{corr/BonizzoniVPPR17,cie/BonizzoniVPPR18ta}.

\begin{table}[!htb]
\centering
\caption{Running times in $\mu$ seconds per input byte.}
\label{t:time}
\begin{tabular}{@{}|l|ccc|ccc|c|@{}}
\hline
Name           & \multicolumn{3}{c|}{\egap} &  \multicolumn{3}{c|}{\bcr} & \egsa \\ 
			   & 1GB     & 8GB   & 32GB  & 1GB    	 & 8GB     & 32GB    & 32GB  \\ \hline
shortreads     & 17.19   & 3.76  & 2.87  & $\times$  & {5.65}  & ~3.96   & 2.08  \\
longreads      & 52.39   & 9.75  & 6.76  & 18.54   	 & {16.01} & 10.88   & 1.89  \\
pacbio.1000    & 24.88   & 3.54  & 1.81  & $\times$  & 54.00   & 36.96   & 1.89  \\
pacbio         & 23.43   & 3.42  & 1.82  & $>70$     & $>50$      & --      & 1.74  \\ \hline
\end{tabular}
\end{table}

Although incomplete, the results show that \bcr\ is competitive for short reads or collections with a large average LCP, while \egap\ clearly dominates in datasets with long reads and relatively small average LCP. In particular, when the available RAM is at least equal to the size of the input, \egap\ can use the semi-external strategy described in Section~\ref{sec:algorithm} and becomes significantly faster. Note that using 32GB RAM both algorithms become much faster: even though they allocate for their use a small fraction of that RAM, the operative system uses the remaining RAM as a buffer and avoids many disk accesses.

\ignore{Space usage: \egap $16$, \bcr $6$, \egsa $54$. 
Peak disk space usage (in bytes per input symbol) for all algorithms.
The numbers include the input, output and temporary files.}

\section{Applications}\label{sec:apps}

In this section we show that the \egap algorithm, in addition to the BWT and LCP arrays, can output additional information that can be used to design efficient {\em external memory} algorithms for three well known problems on sequence collections: the computation of maximal repeats, the all pairs suffix-prefix overlaps, and the construction of succinct de Bruijn graphs. For these problems we describe algorithms which are derived from known (internal memory) algorithms but they process the input data in a single sequential scan. In addition, the amount of RAM used by the algorithms is usually much smaller than the size of inputs since it grows linearly with the number of sequences and the maximum LCP value.

Our starting point is the observation that the \egap algorithm can also output an array which provides, for each $\bwt$ entry, the id of the sequence to which that entry belongs. In information retrieval this is usually called the Document Array, so in the following we will denote it by~$\da$. In Phase~1 the \gSACA algorithm can compute the \da together with the \lcp and \bwt using only additional $4n$ bytes of space to store the \da entries. These partial \da's can be merged in Phase 2 using the $\Zn$ array in the same way as the BWT entries. In the following we use \bwt, \lcp, and \da to denote the multistring BWT, LCP and Document Array of a collection of $m$ sequences of total length $n$. We write $\sxx$ to denote the concatenation $\txx{1} \cdots \txx{m}$ and $\saxx$ to denote the suffix array of $\sxx$. We will use $\sxx$ and $\saxx$ to prove the correctness of our algorithms, but neither $\sxx$ nor $\saxx$ are used in the computations.

\newcommand\entry[2]{{\langle #1, \lcp[#2] \rangle}}
\newcommand\entrz[1]{{\langle #1, \lcp[#1+1] \rangle}}
\newcommand\suffix[1]{\sxx[\saxx[#1], n]}
\newcommand\stack[1]{\mathsf{stack}[#1]}
\newcommand\main{\mathsf{lcpStack}}

\subsection{Computation of Maximal Repeats}\label{sec:MEM}


Different notions of maximal repeats have been used in the bioinformatic literature to model different notions of repetitive structure. We use a notion of maximal repeat from~\cite[Ch.~7]{Gu:97}:  we say that a string $\alpha$ is a {\em Type 1 maximal repeat} if $\alpha$ occurs in the collection at least twice and every extension, i.e. $c\alpha$ or $\alpha c$ with $c\in\A$, occurs fewer times. We consider also a more restrictive notion: we say that a string $\alpha$ is a {\em Type 2 maximal repeat} if $\alpha$ occurs in the collection at least twice and every extension of $\alpha$  occurs at most once.

We first show how to compute Type 1 maximal repeats with a sequential scan of the arrays \bwt and \lcp. The crucial observation is that we have a substring of length $\ell$ that prefixes $\sa$ entries $j, j+1, \ldots, i$ iff $\lcp[h] \geq \ell$ for $h=j+1,\ldots,i$, and both $\lcp[j]$ and $\lcp[i+1]$ are smaller than $\ell$. To ensure that the repeat is maximal, we also require that there exists $h \in [j+1,i]$ such that $\lcp[h]=\ell$ and that $\bwt[j,i]$ contains at least two distinct characters. 

During the scan, we maintain a stack containing pairs $\entry{j}{h}$ such that $j \leq h$; in addition if$\entry{j'}{h'}$ is below $\entry{j}{h}$ on the stack, then $h' < j$. If, when we reach position $i\geq h$, the pair $\entry{j}{h}$ is in the stack this means that all positions $k$ between $j$ and $i$ we have $\lcp[k] \geq \lcp[h]$. To maintain this invariant, when we reach position $i$, if the current top pair $\entry{j}{h}$ has $\lcp[h]<\lcp[i]$, then $\entry{i}{i}$ is pushed on top of the stack. Otherwise, all pairs $\entry{j}{h}$ with $\lcp[h]\geq \lcp[i]$ are popped from the stack; if $\hat{\jmath}$ is the index of the last pair popped from the stack, pair $\entry{\hat{\jmath}}{i}$ is pushed on the stack. The rationale for the latter addition is that for all $\hat{\jmath}\leq j \leq i$ it is $\lcp[j]\geq \lcp[i]$ and therefore the prefix of length $\lcp[i]$ of $\suffix{j}$ is the same as the prefix of the same length of $\suffix{i}$.
It is not difficult to prove that for each stack entry $\entry{j}{h}$, it is $\lcp[j-1]<\lcp[h]$. 



If entry $\entry{j+1}{h}$ is removed from the stack at iteration $i+1$, by the above discussion  $\lcp[j] < \lcp[h]$;  $\lcp[i+1] < \lcp[h]$ (because $\entry{j+1}{h}$ is being removed), and for $k=j+1, \ldots, i$ $\lcp[k]\geq \lcp[h]$. To ensure that we have found a Type 1 maximal repeat we only need to check that $\bwt[j,i]$ contains at least two distinct characters. To efficiently check this latter condition, for each stack entry $\entry{j}{h}$ we maintain a bitvector $b_j$ of size $\Asize$ keeping track of the distinct characters in the array \bwt from position $j-1$ to the next stack entry, or to the last seen position for the entry at the top of the stack.  When $\entry{j}{h}$ is popped from the stack its bitvector is or-ed to the previous stack entry in constant time; if $\entry{j}{h}$ is popped from the stack and immediately replaced with $\entry{j}{i}$ its bitvector survives as it is (essentially because it is associated with an index, not with a stack entry). Clearly, maintaining the bitvector does not increase the asymptotic cost of the algorithm.

To find Type 2 maximal repeats, we are interested in consecutive LCP entries $\lcp[j], \lcp[j+1], \ldots, \lcp[i], \lcp[i+1]$, such that 
$
\lcp[j]< \lcp[j+1] = \lcp[j+2] = \cdots  = \lcp[i] > \lcp[i+1].
$
Indeed, this ensures that for $h=j, \ldots, i$ all suffixes $\suffix{h}$ are prefixed by the same string $\alpha$ of length $\lcp[j+1]$ and every extension $\alpha c$ occurs at most once. If this is the case, then $\alpha$ is a Type 2 maximal repeat if all characters in $\bwt[j,i]$ are distinct since this ensures that also every extension $c\alpha$ occurs at most once. In order to detect this situation, as we scan the \lcp array we maintain a candidate pair $\entry{j+1}{j+1}$ such that $j+1$ is the largest index seen so far for which $\lcp[j]<\lcp[j+1]$. When we establish a candidate at $j+1$ as above, we init a bitvector $b$ marking entries $\bwt[j]$ and $\bwt[j+1]$. As long as the following values $\lcp[j+2], \lcp[j+3], \ldots$ are equal to $\lcp[j+1]$ we go on updating $b$ and if the same position is marked twice we discard $\entry{j+1}{j+1}$. If we reach an index $i+1$ such that $\lcp[i+1]>\lcp[j+1]$, we update the candidate and reinitialize $b$. If we reach $i+1$ such that $\lcp[i+1]<\lcp[j+1]$ and $\entry{j+1}{j+1}$ has not been discarded, then a repeat of Type 2 (with $i-j+1$ repetitions) has been located.

Note that when our algorithms discover Type 1 or Type 2 maximal repeats, we know the repeat length and the number of occurrences, so one can easily filter out non-interesting repeats (too short or too frequent).  In some applications, for example the MUMmer tool~\cite{ploscb/MarcaisDPCSZ18}, one is interested in repeats that occur in at least $r$ distinct sequences, maybe exactly once for each sequence. Since for these applications the number of distinct sequences is relatively small, we can handle this requirements by simply scanning the \da array simultaneously with the \lcp and \bwt arrays and keeping track of the sequences associated to a maximal repeat using a bitvector (or a union-find structure) as we do with characters in the~\bwt.

\ignore{
Our solution for this problem follows closely the one in~\cite{ipl/OhlebuschG10} which in turn was inspired by an earlier Suffix-tree based algorithm~\cite{ipl/GusfieldLS92}. Note that the algorithm in~\cite{ipl/OhlebuschG10} solves the (unrestricted) problem using a Generalized Enhanced Suffix array (consisting of the arrays \saxx, \lcp, and \da) in $\Oh(n+m^2)$ time. Assuming there are $E$ suffix-prefix overlaps longer than~$o$, our algorithm takes $\Oh(n+E)$ time, plus the cost of a single sequential scan of the input arrays.

Another useful information that our algorithm can compute at no extra asymptotic cost is a boolean array we call~\xlcp such that $\xlcp[i-1]=\onex$ iff $\sxx[\saxx[i-1]+\lcp[i]]=\$$, that is, if the $(i-1)$-th lexicographically larger suffix has length $\lcp[i]$ and is therefore a substring of the $i$-th suffix. We call suffixes with this property {\em special} suffixes. To compute the \xlcp array, we maintain a binary array $X$ such that, at the end of iteration $h$ $x[j]=\onex$ iff

We proceed as in Section~\ref{sec:MEM}, maintaining a set of entries $\entry{j}$ such that $\lcp[j]$ is the length of longest common prefix between $\sxx[\sa[j],n]$ and $\sxx[\sa[i],n]$ where $i$ denotes the last scanned position, with the following important differences:
\begin{enumerate}
\item instead of a single stack, here we maintain $m$ distinct stacks, $\stack{1},\ldots,\stack{m}$, one for each input sequence; $\stack{k}$ stores only suffixes belonging to sequence $k$. 
\item we only store entries $\entry{j}$ such that $\lcp[j]$ is special and $\lcp[j]> o$. Hence, when we reach position $j$, we store $\entry{j}$ in $\stack{\da[j]}$ only if $\xlcp[j]=\onex$ and $\lcp[j]> o$.
\end{enumerate}
To maintain the invariant that for all stack entries~$\entry{j}$, $\lcp[j]$ is the length of longest common prefix between $\sxx[\sa[j],n]$ and $\sxx[\sa[i],n]$, when we reach position $i$ we remove all entries such that $\lcp[j]>\lcp[i]$ (regardless of the fact that $\lcp[i]$ is special or greater than~$o$). To this end we maintain an array of lists $\tlist$ such that $\tlist[\ell]$ contains the indexes $k$ such that the entry at the top of $\stack{k}$ has LCP component equal to $\ell$ (this array is a stripped down version of~\cite{ipl/OhlebuschG10}'s  $\mathsf{list}$). In addition, we maintain an additional stack $\main$ containing, in increasing order, the values $\ell$ such that some $\stack{k}$ contains an entry with LCP component equal to~$\ell$.
}

\subsection{All pairs suffix-prefix overlaps}

In this problem we want to compute, for each pair of sequences $\txx{i}$ $\txx{j}$, the longest overlap between a suffix of $\txx{i}$ and a prefix of $\txx{j}$. Our solution follows closely the one in~\cite{ipl/OhlebuschG10} which in turn was inspired by an earlier Suffix-tree based algorithm~\cite{ipl/GusfieldLS92}. The algorithm in~\cite{ipl/OhlebuschG10} solves the problem using a Generalized Enhanced Suffix array (consisting of the arrays \saxx, \lcp, and \da) in $\Oh(n+m^2)$ time, which is optimal since there are $m^2$ overlaps. However, for large collections it is natural to consider the problem of reporting only the overlaps larger than a given threshold~$\tau$ still spending constant time per reported overlap. Our algorithm solves this more challenging problem. 

In the following we say that the suffix starting at $\sa[i]$ is {\em special} iff $\sxx[\sa[i]+\lcp[i+1]]=\stri{\$}$ or, in other words, if the suffix starting at $\sa[i]$ is a substring of the suffix starting at $\sa[i+1]$ (not considering the end-marker $\$$). For example, in Fig.~\ref{fig:BWTetc} (right) the suffixes $\stri{ab}\eosz$, $\stri{abc}\eosone$, $\stri{abcab}\eosz$  are all special. We can modify Phase~2 of our algorithm so that it outputs also a bit array $\xlcp$ such that $\xlcp[i]=\onex$ iff the suffix starting at $\sa[i]$ is special. This modification does not increase the asymptotic cost and requires only $2n$ bits of disk working space (details in the full paper).  

Our algorithm consists of a sequential scan of the arrays \bwt, \lcp, and \da, and \xlcp.
We maintain $m$ distinct stacks, $\stack{1},\ldots,\stack{m}$, one for each input sequence; $\stack{k}$ stores only {\em special} suffixes belonging to sequence~$k$. When the scanning reaches position $j$, we store the pair $\entrz{j}$ in $\stack{\da[j]}$ if and only if $\xlcp[j]=\onex$ and $\lcp[j+1]> \tau$. During the scanning we maintain the invariant that for all stack entries~$\entrz{j}$, $\lcp[j+1]$ is the length of longest common prefix between $\sxx[\sa[j],n]$ and $\sxx[\sa[i],n]$, where $i$ is the next position to be scanned. To this end, when we reach position $i$ we remove all entries $\entrz{j}$ such that $\lcp[j+1]>\lcp[i+1]$. To do this spending constant time for removed entry requires some additional machinery: We maintain an array of lists $\tlist$ such that $\tlist[\ell]$ contains the indexes $k$ for which the entry at the top of $\stack{k}$ has LCP component equal to $\ell$ (this array is a stripped down version of~\cite{ipl/OhlebuschG10}'s  $\mathsf{list}$). In addition, we maintain an additional stack $\main$ containing, in increasing order, the values $\ell$ such that some $\stack{k}$ contains an entry with LCP component equal to~$\ell$.

At iteration $i$, we use $\main$ and the lists in~$\tlist[\cdot]$ to reach all $\stack{k}$ containing entries with LCP component greater than~$\lcp[i+1]$ and we remove them. After the removal, we update $\tlist[\ell]$ where~$\ell$ is the LCP value now at the top of $\stack{k}$. Finally, if $\xlcp[i]=\onex$ and $\lcp[i+1]>\tau$, we add $\entrz{i}$ to $\stack{\da[i]}$; this requires that we also add $\da[i]$ to $\tlist[\lcp[i+1]]$, and that we remove $\da[i]$ from the list $\tlist[\ell]$ where $\ell$ is the previous top LCP value in $\stack{\da[i]}$ (to do this we need to maintain for each element at the top of the stack a pointer to its corresponding \da entry in $\tlist$). Since we perform a constant number of operations per entry, maintaining the above data structures takes $\Oh(n)$ time overall.

The computation of the overlaps is done as in~\cite{ipl/OhlebuschG10}. When the scanning reaches position~$i$, we check whether $\bwt[i]=\$$. If this is the case, then $\sxx[\sa[i],n]$ is prefixed by the whole sequence $\txx{\da[i]}$, hence the longest overlap between a prefix of $\txx{\da[i]}$ and a suffix of $\txx{k}$ is given by the element currently at the top of $\stack{k}$, since by construction these stacks only contain special suffixes whose overlap with $\sxx[\sa[i],n]$ is larger than~$\tau$. To spend time proportional to the number of reported overlaps, instead of accessing all stacks we access only those which are non-empty. This requires that we maintain an additional list containing all values $\ell$ such that $\tlist[\ell]$ is non-empty. For each entry~$\ell$ in this list, $\tlist[\ell]$ gives us the id of the sequences with a suffix-prefix overlap with $\da[i]$ of length~$\ell$.
As in~\cite{ipl/OhlebuschG10}, we have to handle differently the case in which the whole $\txx{\da[i]}$ is a suffix of another sequence, but this can be done without increasing the overall complexity. Since we spend constant time for reported overlap, the overall cost of the algorithm, in addition to the scanning of the \bwt/\lcp/\xlcp/\da arrays, is $\Oh(n+E_\tau)$, where $E_\tau$ is the number of suffix-prefix overlaps greater than~$\tau$. 


\subsection{Construction of succinct de Bruijn graphs}\label{sec:boss}

A recent remarkable application of compressed data structures is the design of efficiently navigable succinct representations of de Bruijn graphs~\cite{latin/BelazzouguiGMPP16,dcc/BoucherBGPS15,wabi/BoweOSS12}. Formally, a de Bruijn graph for a collection of strings consists of a set of vertices representing the distinct $k$-mers appearing in the collection, with a directed edge $(u,v)$ iff there exists a $(k+1)$-mer $\alpha$ in the collection such that $\alpha[1,k]$ is the $k$-mer associated to $u$ and $\alpha[2,k+1]$ is the $k$-mer associated to $v$.

The starting point of all de Bruijn graphs succinct representation is the BOSS representation~\cite{wabi/BoweOSS12}, so called from the authors' initials. For simplicity we now describe the BOSS representation of a $k$-order de Bruijn graph using the lexicographic order of $k$-mers, instead of the co-lexicographic order as in~\cite{wabi/BoweOSS12}, which means we are building the graph with the direction of the arcs reversed. This is not a limitation since arcs can be traversed in both directions (or we can apply our construction to the input sequences reversed).

Consider the $N$ $k$-mers appearing in the collection sorted in lexicographic order. For each $k$-mer $\alpha_i$ consider the array $C_i$ of distinct characters $c\in \A \cup\{\$\} $ such that $c\alpha_i$ appears in the collection. The concatenation $\Wa = C_1 C_2 \cdots C_N$ is the first component of the BOSS representation. The second component is a binary array $\last$, with $|\last| = |\Wa|$, such that $\last[j]=\onex$ iff $\Wa[j]$ is the last entry of some array $C_i$. Clearly, there is a bijection between entries in $\Wa$ and graph edges; in the array $\last$ each sequence $\zerox^{i}\onex$ ($i\geq 0$) corresponds to the outgoing edges of a single vertex with outdegree $i+1$. Finally, the third component is a binary array $\Wx$, with $|\Wx| = |\Wa|$, such that $\Wx[j]=\onex$ iff $\Wa[j]$ comes from the array $C_i$, where $\alpha_i$ is the lexicographically smallest $k$-mer prefixed by $\alpha_i[1,k-1]$ and preceded by $W[j]$ in the collection. Informally, this means that $\alpha_i$ is the lexicographically smallest $k$-mer with an outgoing edge reaching $W[j]\alpha_i[1,k-1]$. Note that the number of $\onex$'s in $\last$ and $\Wx$ is exactly $N$, i.e. the number of nodes in the de Bruijn graph. 

We now show how to compute $\Wa$, $\last$ and $\Wx$ by a sequential scan of the $\bwt$ and $\lcp$ array. The crucial observation is that the suffix array range prefixed by the same $k$-mer $\alpha_i$ is identified by a range $[b_i,e_i]$ of LCP values satisfying
$\lcp[b_i]<k$, $\lcp[\ell] \geq k$ for $\ell=b_i+1,\ldots,e_i$ and $\lcp[e_i+1]<k$. Since $k$-mers are scanned in lexicographic order, by keeping track of the corresponding characters in the array $\bwt[b_i,e_i]$ we can build the array $C_i$ and consequently $\Wa$ and $\last$. To compute $\Wx$ we simply need to keep track also of suffix array ranges corresponding to $(k-1)$-mers. Every time we set an entry $\Wa[j]=c$ we set $\Wx[j]=\onex$ iff this is the first occurrence of $c$ in the range corresponding to the current $(k-1)$-mers. 

If, in addition to the \bwt and \lcp arrays, we also scan the \da\ array, then we can keep track of which sequences contain any given graph edge and therefore obtain a succinct representation of the colored de Bruijn graph~\cite{bioinformatics/MuggliBNMBRGPB17etal}. Finally, we observe that if our only objective is to build the $k$-order de Bruijn graph, then we can stop the phase 2 of our algorithm after the $k$-th iteration. Indeed, we do not need to compute the exact values of LCP entries greater than $k$, and also we do not need the exact BWT but only the BWT characters sorted by their length~$k$ context.



\end{document}